\documentstyle[aps,psfig,preprint]{revtex}
\newcommand{\lan}{\langle}
\newcommand{\ran}{\rangle}

\begin{document}
\title{The jamming transition of Granular Media}

\author{Antonio Coniglio$^{a}$ and Mario Nicodemi$^{b}$}
\address{
\vspace{0.2cm}
$a)$ Dipartimento di Scienze Fisiche, Sezione INFM and INFN di Napoli, \\
Universit\'a di Napoli ``Federico II'', 
Mostra d'Oltremare, Pad. 19, 80125, Napoli, Italy $~~~$ coniglio@na.infn.it \\
$b)$ Imperial College of Science, Technology and Medicine, 
 Department of Mathematics \\
 Huxley Building, 180 Queen's Gate, London, SW7 2BZ, U.K. 
$~~~$ nicodemi@ic.ac.uk \\
}

\maketitle

\begin{abstract}
A statistical mechanical approach to granular material is proposed.
Using lattice models from standard Statistical Mechanics and 
results from a mean field replica approach we find a 
jamming transition in granular media closely related to the glass transition 
in super-cooled liquids. 
These models reproduce the logarithmic relaxation in granular 
compaction and reversible-irreversible lines, in agreement with
experimental data. The models also  exhibit aging effects and breakdown 
of the usual fluctuation dissipation relation. 
It is shown that the glass transition may be 
responsible for the logarithmic relaxation and may be related to the 
cooperative effects underlying many phenomena of granular materials 
such as the Reynolds transition.
\end{abstract}

\section{Introduction}
Despite their importance for industrial applications, 
non-thermal disordered systems as granular media 
have just recently begun to be systematically studied in the physics 
community \cite{JNBHM}.
In particular, concepts from Statistical Mechanics seems to be 
successful to describe these systems as suggested 
in his pioneering work by Edwards \cite{Edwards}. 
Actually, granular media are composed 
by a large number of single grains and, as much as systems of standard  
Statistical Mechanics, each of their macroscopic states corresponds 
to a huge number of microstates. 
Furthermore, they show very general reproducible macroscopic behaviours 
whose general properties are not material specific and 
which are statistically characterised by very few control parameters, 
such as their density, load or amplitude of external drive \cite{JNBHM}. 
Granular media are ``non-thermal'' systems 
since thermal energy plays no role with respect, for instance, to 
gravitational energies involved in grain displacements, however, 
thermal motion can be replaced by agitation 
induced by shaking or other driving and this allows the system 
to explore its space of configurations.

A ``tapping'' dynamics has several control parameters. 
A very important one, related to the amplitude of vibrations of 
the external driving, has appeared to be the adimensional ratio 
$\Gamma_{ex}=a/g$ \cite{JNBHM,Knight}, where $a$ is the shake peak 
acceleration and $g$ the gravity acceleration constant. 
It has been suggested that $\Gamma_{ex}$ plays a role very similar 
to ``temperature'' in thermal systems. For simplicity, below we suppose 
the characteristic frequency, $\omega$, of external 
driving is fixed, and actually 
in typical experiments as those in Ref.s \cite{Knight,Novak} 
we are interested in, it plays a minor role with respect to 
$\Gamma_{ex}$ (see \cite{JNBHM}). 
Material parameters such as restitution, friction coefficients, 
presence of moisture, etc., are also to be usually considered, but they 
generally do not affect the overall behaviours. 

Experimentally it is found that at low $\Gamma_{ex}$ relaxation processes as 
density compaction under tapping are logarithmically slow.
An essential ingredient in the dynamics of dense granular media is, instead, 
the presence of ``frustrating'' mechanisms on the motion of grains
due to steric hindrance and hard core repulsion. 
We have introduced Frustrated Lattice Gases to describe the dynamics 
of granular materials in the regime of high densities or 
not too strong shaking \cite{CH,NCH,Caglioti} . 
They interestingly allow to interpret, 
in a single framework, several different 
properties of granular media ranging from 
logarithmic compaction, density fluctuations, 
``irreversible-reversible" cycles, 
aging, breakdown of fluctuation dissipation relation, segregation, 
to avalanches effects, the Reynolds transition and several others
\cite{NCH,Caglioti,NC_aging,N_fdt,NC_mspowders}. 
In particular they predict the existence of a definite 
``jamming'' transition in granular media, in strict correspondence with 
the glass transition of glass formers and spin glasses 
\cite{JNBHM,angell,BCKM,BY}. 
Furthermore, they are new important applications 
of the statistical mechanics to powders\cite{Edwards,mehta1}. 

\section{Frustrated Lattice Gas models for granular media}

The very simple schematic models for gently shaken 
granular media we introduced are based on a drastic reduction of 
the degrees of freedom of the systems to those we suppose are essential. 
The models consist of a system of elongated grains which move on a lattice. 
Grains are subject to gravity and, eventually, to shaking, which is 
simulated with a driven diffusion like Monte Carlo dynamics. 
The crucial ingredient of the models is the presence of 
geometric constraints on the motion of the grains, thus, 
they have been called Frustrated Lattice Gases \cite{CH,NCH}.
Particles in our models occupy the 
sites of, say, a square (or cubic) lattice 
(see Fig.\ref{lattice}). They have also an internal degree of
freedom $S_i =\pm 1$ corresponding to their possible ``orientations'' 
along the two lattice axes. 
Two nearest neighbour sites can be both occupied only if the particles do 
not overlap.

Each ``experimental'' 
tap can be divided in two parts: one where the grains
average kinetic energy is finite and a second where it goes to zero.
Thus, in our models grains undergo a schematic driven 
diffusive Monte Carlo (MC) dynamics: in absence of vibrations they are 
subject only to gravity and they can move just downwards always fulfilling 
the non overlap  condition; 
the presence of vibration is introduced by allowing the particles to
diffuse with a probability $p_{up}$ to move upwards and 
a probability $p_{down} =1-p_{up}$ to move downwards. The quantity 
$-\ln (x_0)/2$, with $x_0=p_{up}/p_{down}$, as we will see, plays  
the role of an effective temperature and can be related to the 
experimental tap vibrations intensity, $\Gamma_{ex}$. 

\subsection{Hamiltonian description}

\begin{figure}
\centerline{\psfig{figure=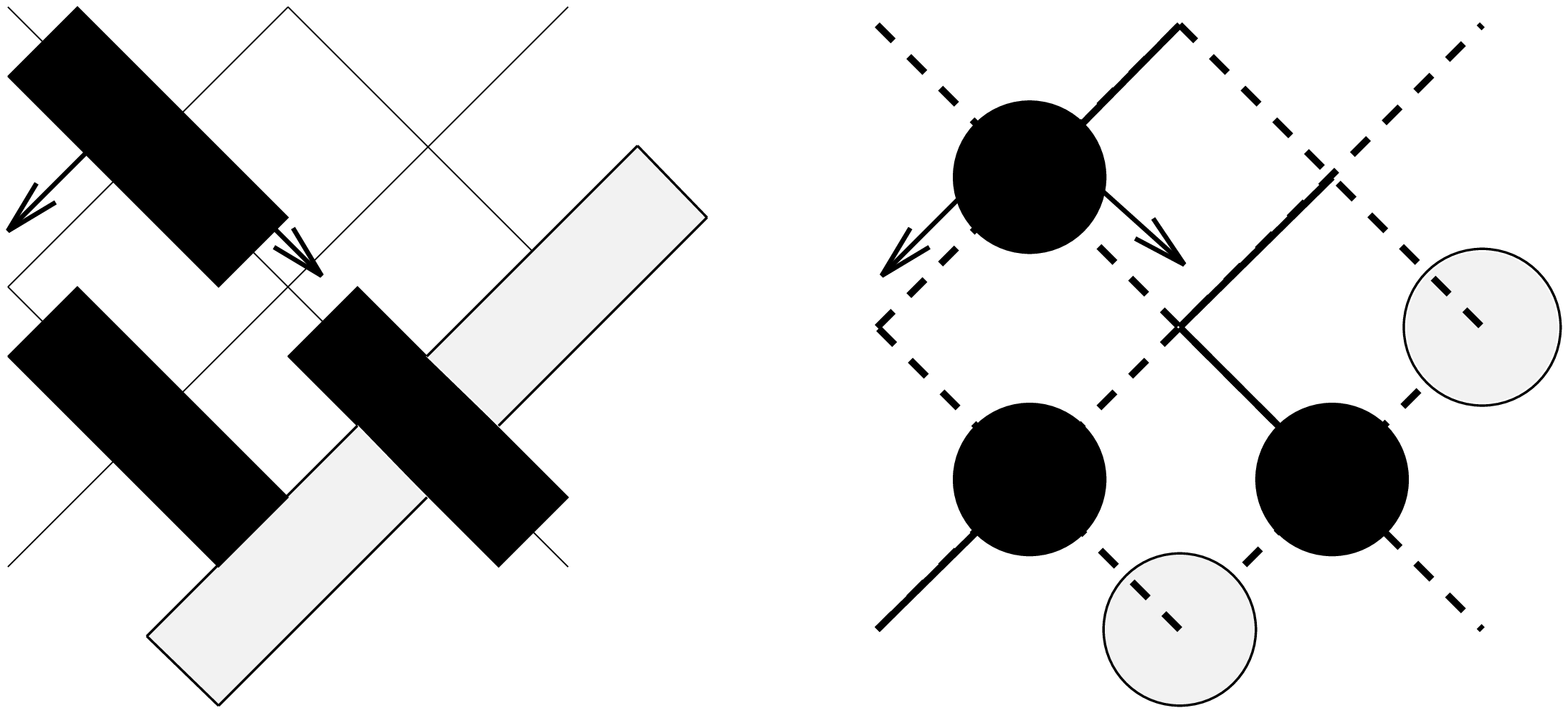,height=6.0cm,angle=-90}}
\vspace{0.5cm}
\caption{A schematic picture of 
the two kind of frustrated lattice gas models described in the 
text. {\em Left:} the Tetris model. 
{\em Right:} the Ising Frustrated Lattice Gas, IFLG.
Straight and dashed lines represent the two kind of interactions
$\epsilon_{ij}=\pm 1$. 
Filled circles are present particles with ``orientation" 
$S_i=\pm 1$ (black/white).
} 
\label{lattice}
\end{figure}

The general model introduced above can be described in terms of a 
standard lattice gas of Statistical Mechanics \cite{CH,NCH}. 
The system Hamiltonian must have an hard core repulsion term 
($J\rightarrow\infty$): 
\begin{equation}
H_{HC}=J\sum_{\langle ij\rangle } f_{ij}(S_i,S_j)n_i n_j ~~ .
\label{H}
\end{equation}
Here $n_i=0,1$ are occupancy variables describing the positions of grains,
$S_i=\pm 1$ are ``spin'' variables associated to the orientations of the
particles, $J$ represents the infinite repulsion
felt by the particles when they have the wrong orientations. 
The hard core repulsion function 
$f_{ij}(S_i,S_j)$ is 0 or 1 depending whether the configuration 
$S_i,S_j$ is right (allowed) or wrong (not allowed), see Fig.\ref{lattice}.

The choice of $f_{ij}(S_i,S_j)$ depends on the particular model. 
In particular, here we consider two models: the Tetris and the 
Ising Frustrated Lattice Gas (IFLG). 
The Tetris model is made of elongated particles 
(see Fig.~\ref{lattice}), 
which may point in two 
(orthogonal) directions coinciding with the two lattice bond orientations. 
In this case $f_{ij}(S_i,S_j)$ is given by \cite{Caglioti}
$f^{Tetris}_{ij}(S_i,S_j)=1/2(S_i S_j -\epsilon_{ij} (S_i+ S_j) +1)$
here $\epsilon_{ij}=+ 1 $ for bonds along one direction of the lattice and 
$\epsilon_{ij}=- 1 $ for bonds on the other. 
In order to have a non trivial behaviour 
the dynamics of the Tetris has a crucial {\em purely kinetic constraint}: 
particles can flip their ``spin'' only if three of their own 
neighbours are empty. 

A real granular system may contain more disorder 
due to the presence of a wider grain 
shape distribution or to the absence of a regular underlying lattice. 
Typically, each grain moves in the disordered environment generated by its 
neighbours. In order to 
schematically consider these effects within the above context, 
an other kind of ``frustrated lattice gas'' was introduced, 
made of grains moving in a lattice with quenched geometric disorder.  
Such a model, the Ising Frustrated Lattice Gas (IFLG), 
has the following hard core repulsion function, $f_{ij}(S_i,S_j)$ \cite{NCH}:
$f_{ij}^{IFLG}(S_i,S_j)=1/2(\epsilon_{ij} S_i S_j -1)$
where $\epsilon_{ij}=\pm 1$ are quenched random interactions associated to the
edges of the lattice, describing the fact that particles must satisfy the 
geometric constraint of the environment considered as ``practically'' 
quenched (see Fig.\ref{lattice}). 
The IFLG shows a non trivial dynamics {\em without} 
the necessity to introduce kinetic constraints. 

The phase diagram of the Tetris Hamiltonian corresponds to the usual 
antiferromagnetic Ising model with dilution. 
The Hamiltonian of the IFLG exhibits richer behaviours. In the limit where 
all sites are occupied ($n_i=1$ $\forall i$), it becomes equal 
to the usual $\pm J$ Ising Edwards-Anderson Spin Glass \cite{BY}. 
In the limit $J\rightarrow\infty$ (which we consider below), 
a version of Site Frustrated Percolation is recovered 
\cite{Coniglio,NC}.

The other important contribution to the full Hamiltonian of a granular pack 
we consider below must be gravitational energy: ${\cal H}=H_{HC}+H_G$, 
where 
$H_G=g\sum_i n_iy_i$, 
and $g$ is the gravity constant and $y_i$ is the height of particle $i$
(grains mass and lattice spacing are set to unity). 
The temperature, $T$, of the present Hamiltonian system (with $J=\infty$)
is related to the ratio $x_0=p_{up}/p_{down}$ via the following relation: 
$e^{-2g/T} = x_0$. It is useful to define the adimensional quantity 
$\Gamma\equiv 1/\ln(x_0^{-1/2})= T/g$, which we assume 
plays the same role as the amplitude of the vibrations 
in real granular matter, 
that is $\Gamma$ is a smooth function of $\Gamma_{ex}$.

It is easy to show\cite{NC_mspowders} that if the system reaches equilibrium, 
the temperature $T$ is:
\begin{equation}
T^{-1}=\frac{\partial \ln\Omega}{\partial E}
\end{equation}
where $\Omega$ is the number of configurations corresponding to the 
gravitational energy E. 
Note that for samples with constant particle density, $E$ is proportional 
to the volume and thus $1/T$ coincides with Edwards' compactivity
\cite{Edwards} apart from a proportionality constant.

\section{The dynamics of compaction} 

\begin{figure}[h]
\vspace{-1cm}
\hspace{-3cm}\centerline{\psfig{figure=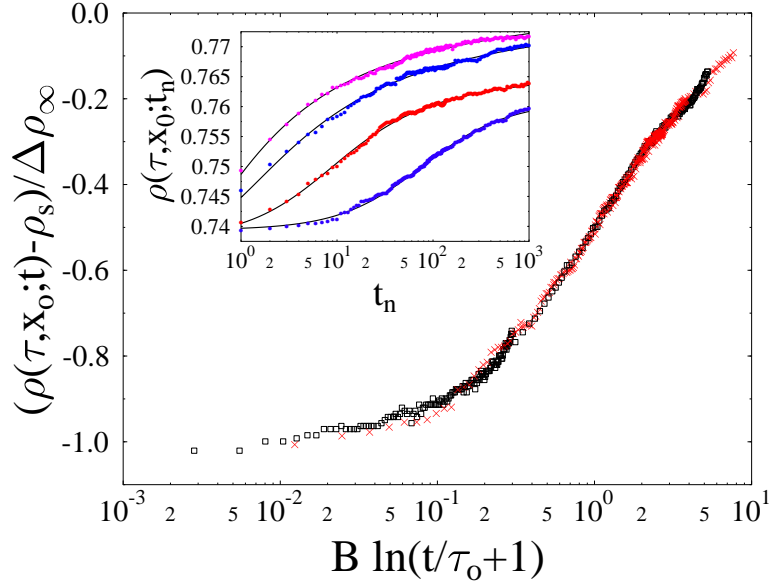,height=12cm,angle=-90}}
\vspace{-1.5cm}
\caption{
The compaction of granular media at low shaking amplitudes. 
Experimental data from Knight et al. 
(squares) and MC data about IFLG (circles) 
rescaled according eq.~(\ref{logrel_kn}).
Insert: density $\rho(\tau,x_0;t_n)$ from MC data 
as a function of tap number $t_n$, 
for tap amplitudes $x_0=0.001,0.01,0.05,0.1$ (from bottom to
top) and duration $\tau=3.28\cdot 10^1$. 
The superimposed curves are logarithmic fits from 
eq.~(\ref{logrel_kn}). 
}
\label{den_tap}
\end{figure}

To describe experimental observations 
about grain density relaxation under a sequence of taps  
a logarithmic law was proposed in Ref.~\cite{Knight}:
\begin{equation}
\rho(t_n)=\rho_{\infty}-\Delta\rho_{\infty}/[1+B ~ \ln(t_n/\tau_1+1)]
\label{logrel_kn}
\end{equation}
This law has proved to be satisfied very well by relaxation data in
the IFLG model \cite{NCH}, which can be excellently rescaled with 
experimental data. In Fig.\ref{den_tap} MC compaction data 
for four different amplitudes, $\Gamma$, as well as the
experimental data for three different amplitudes, $\Gamma_{ex}$, 
are collapsed on a single curve using eq.(\ref{logrel_kn}).
The agreement is very satisfactory 
(details about these data are in Ref.\cite{NCH}
).

Interestingly, the results from Tetris are similar \cite{Caglioti}, 
but the asymptotic density $\rho_{\infty}$ is numerically 
indistinguishable from 1, thus almost independent on $x_0$, a fact in 
contrast with both IFLG and experimental results  
\cite{Knight}.

\section{Glassy behaviours of granular media}

We have seen that compaction shows extremely long relaxation times.
Actually we show that granular systems are typically in off equilibrium 
configurations and that they may undergo a ``jamming'' transition.
In particular, in the same framework of the above models, 
we show how experimental results about ``memory'' effects  
(the so called ``irreversible-reversible'' cycles) 
outline the existence of a dynamical glassy transition, which can be defined 
in a similar way to the glassy transition in real glass formers. 

It has been shown in mean field approximation \cite{ANS} and
numerically for finite dimensional systems \cite{NC}, 
that the IFLG, in absence of gravity, 
exhibits a spin glass (SG) transition at high
density (or low temperature) similar to the one found in the 
p-spin model. 
J.J. Arenzon has extended the mean field solution of the IFLG in presence 
of gravity \cite{Arenzon}, showing that at low $\Gamma$ the system 
is frozen in a SG-like phase, but at higher $\Gamma$ it separates 
in a frozen SG phase at the bottom and a fluid phase on top. Above 
a critical $\Gamma$ the system is entirely fluid \cite{mehta2}. 

\subsection{``Irreversible-reversible'' cycles and the dynamical 
glass transition}

\begin{figure}[ht]
\vspace{-1cm}
\centerline{\psfig{figure=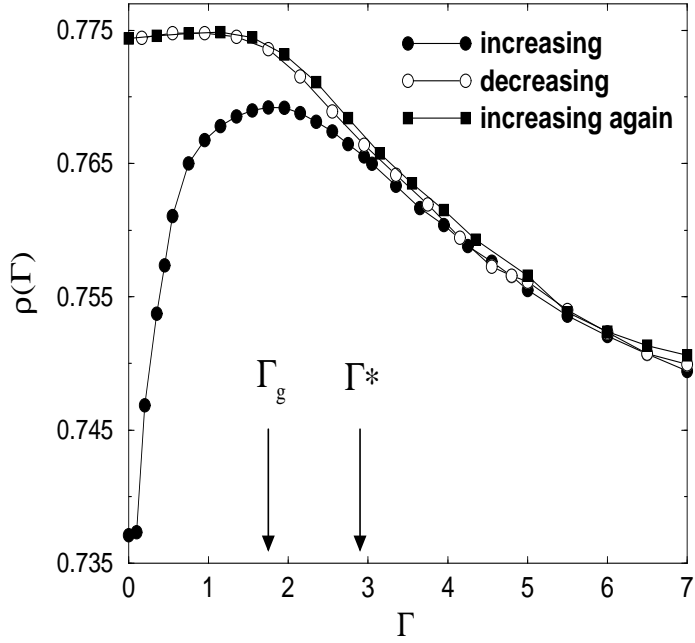,height=10cm,width=10cm,angle=-90}}
\vspace{0.5cm}
\caption{The static bulk density, $\rho(\Gamma)$, of the IFLG model 
as a function of 
the vibration amplitude, $\Gamma$, in cyclic vibration sequences. 
The system is shaken with an amplitude $\Gamma$ which at first 
is increased (filled circles), then is decreased (empty circles) 
and, finally, increased again (filled squares) with a
given  ``annealing-cooling" velocity 
$\gamma\equiv \Delta\Gamma/\tau$. Here we fixed $\gamma=1.25~10^{-3}$. 
The upper part of the cycle is approximately ``reversible" 
(i.e., empty circles and filled squares fall roughly on the 
same curve). 
The data compare rather well with the experimental data of 
Novak et al.. 
$\Gamma^*$ is approximately the point where the ``irreversible" and 
the ``reversible" branches meet. $\Gamma_g$ signals the 
location of a ``jamming transition".
} 
\label{den_hys}
\end{figure}

Tapping experiments typically show ``irreversible-reversible" cycles 
\cite{Novak} (see Fig.\ref{den_hys}): during a sequence of taps, if 
the system is successively shaken at increasing vibration amplitudes, 
its bulk density typically grows, as in compaction, and then, 
after a characteristic $\Gamma_{ex}$ starts decreasing. However, if 
the amplitude of shaking is decreased back to zero, 
the density generally doesn't follow the same path 
since it keeps growing. 
In this sense these observations outlines the existence of ``memory effects''.

\begin{figure}[ht]
\vspace{-2cm}
\centerline{\psfig{figure=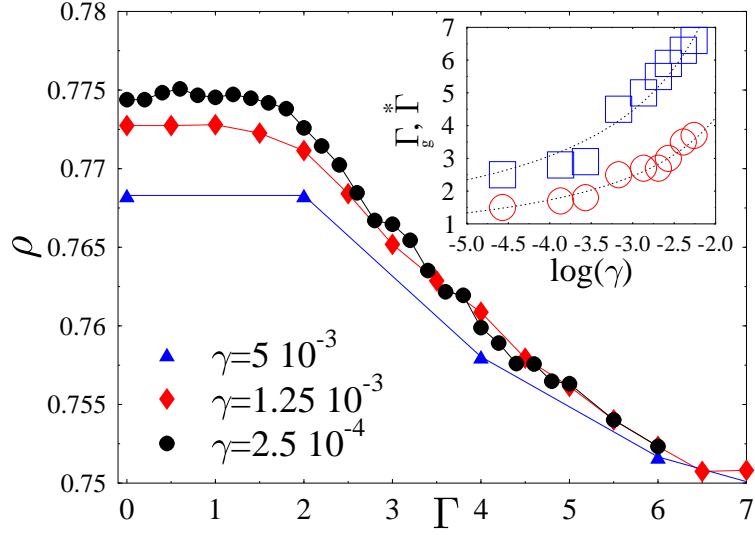,height=12cm,angle=-90}}
\vspace{-1.5cm}
\caption{
{\em Main frame:} As in Fig.\ref{den_hys}, 
the density, $\rho(\Gamma)$, is plotted as a function of 
the vibration amplitude, $\Gamma$ 
for three values of the ``cooling'' velocity $\gamma$.
Here only the descending parts of the cycles are shown. As in 
glasses, a too fast cooling drives the system out of 
equilibrium. The position of the shoulder, $\Gamma_g(\gamma)$, 
schematically individuates a ``jamming transition". 
{\em Inset:} numerical estimate, in the IFLG model of the 
dependence of $\Gamma_g(\gamma)$ (circles) and $\Gamma^*(\gamma)$ (squares) 
on the cooling rate $\gamma$. Superimposed are logarithmic fits 
in analogy to those for 
the glass transition temperature, $T_g(\gamma)$, in glasses [4]. 
When $\gamma\rightarrow 0$ we roughly find 
$\Gamma_g(0) =  \Gamma^*(0)$. 
} 
\label{den_hys_up_branch}
\end{figure}

In analogy to real experiments, cycles of taps were performed in the IFLG 
in which the vibration amplitude 
$\Gamma$ was varied in sequences of increments 
$\gamma=\Delta \Gamma/\tau$, with constant taps duration $\tau$. 
After each tap the static bulk density of the system 
$\rho(\Gamma_n)$ ($n$ is the $n$-th tap number) was then measured. 
The data are qualitatively very similar to those reported in real 
experiments on dry granular packs \cite{Novak}. Furthermore, 
they allow to define a ``jamming'' transition 
point $\Gamma_g(\gamma)$ in analogy to the glass transition in 
glass formers \cite{NC_aging}, 
as explained in Fig.\ref{den_hys_up_branch}.

\subsection{Two-times correlation functions}

Granular media in the above 
dynamical situations show ``aging'' too \cite{NC_aging}. 
To see this, as much as in glassy systems, 
it is useful to consider two time correlation functions as ($t\ge t'$): 
$C(t,t')=
[\lan\rho(t)\rho(t')\ran-\lan\rho(t)\ran\lan\rho(t')\ran]/
[\lan\rho(t')^2\ran-\lan\rho(t')\ran^2]
$
where $\rho(t)$ is the bulk density of the system at time $t$. 

\begin{figure}[ht]
\vspace{-2cm}
\hspace{-2cm}
\centerline{\psfig{figure=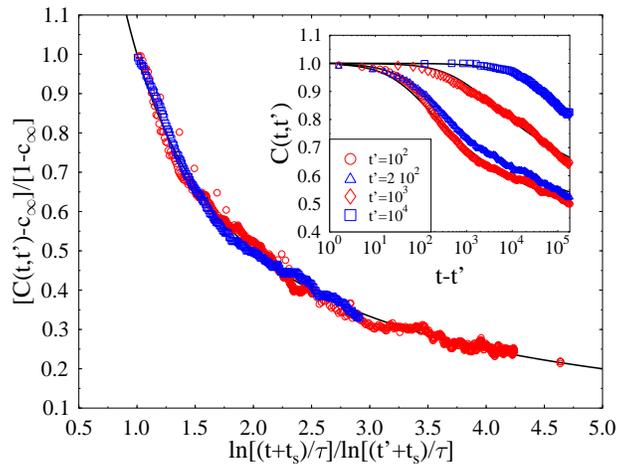,height=10cm,angle=-90}}
\vspace{-1.5cm}
\caption{ The two time density-density correlation function, 
$(C(t,t')-c_{\infty})/(1-c_{\infty})$, 
as a function of the scaling variable 
$\alpha=\ln[(t+t_s)/\tau]/\ln[(t'+t_s)/\tau]$. 
Scaled on the same master function are data from both models considered 
in the present paper (Tetris, squares, and IFLG, circles) for 
$\Gamma=-1/\ln(x_0)$ with $x_0\in[10^{-4},10^{-1}]$. 
{\em Inset:} The correlation $C(t,t')$ for the Tetris at $\Gamma=0.22$ 
(or $x_0=0.01$) as a function of $t-t'$ for $t'=10^2, 2~10^2, 10^3, 10^4$. 
} 
\label{fig_cor_sca}
\end{figure}

The fit, at low $\Gamma$, of the two time correlation function, $C(t,t')$, 
(on five decades in MC time), reveals the following interesting 
approximate scaling form:
\begin{equation}
C(t,t')=(1-c_{\infty}) \frac{\ln[(t'+t_s)/\tau]}{\ln[(t+t_s)/\tau]} 
+ c_{\infty}
\label{cor_sca}
\end{equation}
where $\tau$ (which now is not the ``tap duration'' as before), 
$t_s$ and $c_{\infty}$ are fit parameters. 
Very interesting is the fact that the above behaviour is found in both 
the discussed models (Tetris and IFLG) \cite{NC_aging}. 
The data for the two models, for several values of $\Gamma$, 
rescaled on a single universal master 
function, are plotted in Fig.~\ref{fig_cor_sca}. 
It is interesting the existence of such a scaling behaviours 
in off equilibrium dynamics of apparently different systems \cite{CN_sca}.

\subsection{Response functions and the 
``fluctuation-dissipation'' relation}

In Ref. \cite{N_fdt} it was argued that in granular media it is possible 
to formulate a link between response and fluctuations, i.e., the analogous 
of a ``fluctuation dissipation theorem" (FDT) (see \cite{Kubo}), 
where the amplitude of external vibrations plays the role of the 
usual ``temperature''. In typical situations the FDT coincides neither 
with its usual version at equilibrium nor with the extensions 
valid in off equilibrium thermal systems in the so called ``small 
entropy production" limit \cite{CDK}. 
The origin of universalities in off equilibrium dynamics 
in the above heterogeneous classes of materials is still unanswered. 

Our systems are  ``shaken'' at a given 
amplitude $x_0$ and the average grains height, 
$h_0(t)=\lan H(t)\ran$ (with $H(t)=\sum_iy_i(t)/N$ ),
recorded as long as the ``mean square displacement"  
$B(t,t')\equiv \lan [H(t)-H(t')]^2 \ran$.
To measure the response function, 
the average height, $h_1(t,t_w)$, was also recorded 
in an identical copy of the system (a ``replica'') 
perturbed by a small increase of the shaking amplitude, after a time $t_w$. 

The difference in heights between the perturbed and unperturbed systems 
$\Delta h(t,t_w)=h_1(t,t_w)-h_0(t)$ 
is by definition the integrated response. 
FDT or its generalisations concern the relation between 
$\Delta h(t,t_w)$ and the displacement, $B(t,t_w)$. 
In analogy to thermal systems \cite{BCKM}, 
in Ref.\cite{N_fdt} was proposed that:
\begin{equation}
\Delta h(t,t_w) \simeq \frac{X}{2} \Delta(\Gamma^{-1}) B(t,t_w)
\label{FDT}
\end{equation}
Eq.(\ref{FDT}) states that, after transients, in a granular system the 
measure of the height variation after a change in shaking amplitude, 
should be proportional to the ``displacement" recorded during an 
unperturbed run. 
In the simplest situations, as equilibrium thermal systems, the 
proportionality factor, $X$, is a constant, 
but more generally, $X$ is function of $t_w$ and $t$ themselves. 
Interestingly, in glassy systems the quantity 
$g\Gamma/X$ has the meaning of an ``effective temperature" \cite{CKP,K_rheo}. 

\begin{figure}[ht]
\centerline{\psfig{figure=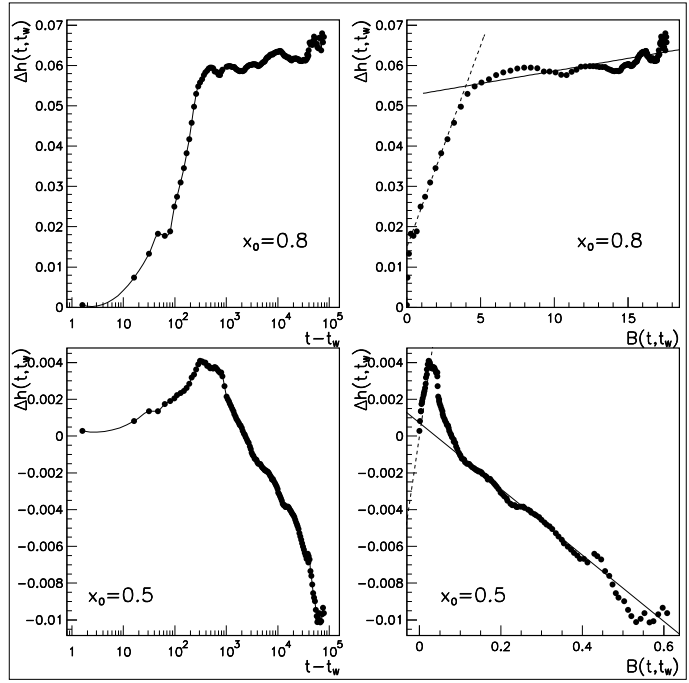,height=9cm,width=9cm,angle=0}}
\vspace{0.5cm}
\caption{
The figures on the left show, 
as a function of $t-t_w$, the average height difference, 
$\Delta h(t,t_w)\equiv h_1(t,t_w)-h_0(t)$, 
of a reference system shaken at a given $x_0$ and a replica perturbed 
after $t_w$ by shaking at $x_0+\Delta x_0$ ($\Delta x_0=.002$). 
On the right, a check of the generalized 
fluctuation dissipation relation of eq.~(\ref{FDT}) is given. 
The integrated response, $\Delta h(t,t_w)$, is plotted 
as a function of the displacement of the reference system, $B(t,t_w)$. 
Systems are shaken at different ``amplitudes'' $x_0$ 
($x_0=0.8$ top, $x_0=0.5$ bottom), with replicas perturbed after a $t_w=370$. 
At low $x_0$, negative responses appear, but 
in agreement with eq.~(\ref{FDT}), $\Delta h$ is asymptotically still 
approximately linear in $B$. 
}            
\label{ut_all}
\end{figure}

In the present model, eq.~(\ref{FDT}) seems to be approximately valid, 
also if in typical off-equilibrium situations $X$ slowly depends on $\Gamma$ 
and $t_w$. This is shown, for high $x_0$, in the top panels 
of Fig.\ref{ut_all}: in the long time regime $X$ is equal to 1, 
showing that, in the high ``temperature" and 
low density region, the usual equilibrium version of FDT is obeyed. 
In the low $x_0$ region, the above picture changes, as shown in  
the bottom panels Fig.\ref{ut_all}: after an early transient
the response, $\Delta h$, is {\em negative} and, thus, 
$X$ is negative, which would asymptotically correspond to a negative 
``effective temperature".

\section{Conclusions}

In conclusion, the present paper has dealt with 
the understanding of the ``jamming'' transition in granular media 
via the introduction of 
models from standard statistical mechanics to describe these ``non-thermal'' 
systems (the IFLG and Tetris, \cite{NCH,Caglioti}). 

They exhibit a logarithmic compaction when subject to gentle shaking in 
presence of gravity, a compaction extremely close to what is experimentally 
observed in granular packs \cite{Knight}. The presence of such slow dynamics 
is linked to the existence of a ``jamming" transition in granular media. 
In facts, in the high density region, at $\rho_m$, the models undergo a 
structural arrest where grains self-diffusivity becomes zero \cite{NCH}.

Self-diffusion suppression at $\rho_m$ signals that, above such a density, 
it is impossible to obtain a macroscopic rearrangement of grains without 
increasing the system volume, a fact interestingly similar to the phenomenon 
of the Reynolds dilatancy transition \cite{NCH}. 

A very important fact is that results from the present models 
are in excellent agreement with known experimental ones, and their 
many new outcomes must be still experimentally investigated. 
We have also discussed the intriguing connections of granular media with 
others materials, as glassy systems or spin glasses, 
where geometrical disorder and frustration play a crucial role. 

\section{Acknowledgements}   
We thank INFM-CINECA for CPU time on Cray-T3D/E.
Work partially supported by the TMR Network ERBFMRXCT980183,
MURST-PRIN 97 and INFM-PRA 99.

\pagebreak

\pagebreak

\pagebreak

\pagebreak

\pagebreak

\pagebreak

\end{document}